# Fabrication and Characterization of Poly(methyl methacrylate)/CaCu$_3$Ti$_4$O$_{12}$ Composites.

P.Thomas,[a] R.S. Ernest Ravindran.[a] K.B.R.Varma.[b●]

[a] Dielectric Materials Division, Central Power Research Institute, Bangalore : 560 080, India

[b] Materials Research Centre, Indian Institute of Science, Bangalore: 560012,India

## Abstract

Composites comprising Poly(Methyl Methacrylate) (PMMA) and CaCu$_3$Ti$_4$O$_{12}$ (CCTO) via melt mixing followed by hot pressing were fabricated. These were characterized using X-ray diffraction (XRD), thermo gravimetric (TGA), scanning electron microscopy (SEM) and Impedance analyser for their structural, morphology and dielectric properties. Composites were found to have better thermal stability than that of pure PMMA. The composite, with 38 Vol % of CCTO (in PMMA), exhibited remarkably low dielectric loss at high frequencies and the low frequency relaxation is attributed to the space charge polarization/MWS effect. Theoretical models were employed to rationalise the dielectric behaviour of these composites. At higher temperatures, the relaxation peak shifts to higher frequencies, due to the merging of both β and α relaxations into a single dielectric dispersion peak. The AC conductivity in the high frequency region was attributed to the electronic polarization.

## INTRODUCTION

In recent years, particulate composites (comprising ceramic crystallites) possessing high permittivity have become increasingly important for high energy density capacitor applications. The permittivity values obtained in these composites are invariably low as compared to that of the ceramics, because of the low permittivity associated with polymer matrices. Though the bulk ceramics possess high permittivity (>1000), the incorporation of highest possible volume fraction of these crystallites into the polymer matrix, the effective permittivity that could be achieved is lower than that of ceramics [1-8]. The most widely used high permittivity ceramic-polymer composites are those based on ferroelectric BaTiO$_3$(BT),

---
[●] Corresponding author : Tel. +91-80-2293-2914; Fax: +91-80-2360-0683.
E-mail : kbrvarma@mrc.iisc.ernet.in (K.B.R.Varma)

BaSrTiO$_3$(BST),Pb(Mg$_{1/3}$Nb$_{2/3}$)O$_3$-PbTiO$_3$(PMNPT), and Pb(Zr,Ti)O$_3$(PZT) [2-8]. For capacitor applications, the high $\varepsilon_r^{'}$ of ferroelectric materials accompanied by low loss is highly desirable. For practical applications, however, it is also necessary for this high $\varepsilon_r^{'}$ to be maintained over a wide temperature range. Hence there is a continued interest in the search of the materials having high permittivity ($\varepsilon_r^{'}$) over a wide temperature range [8]. Recently, the CaCu$_3$Ti$_4$O$_{12}$ (CCTO) ceramic has gained considerable attention due to its large permittivity ($\varepsilon$~10$^{4-5}$) [9] and has been used as a filler and studied [10-17] to explore the possibility of obtaining a new generation composites associated with high permittivity for capacitor applications in electrical circuits. The permittivity was reported to increase as the CCTO content increased in the polymer matrix and decreased as the frequency increased [10-17] and the permittivity as high as 740 at 1 kHz was achieved [10].

The Poly (Methyl Methacrylate) (PMMA), a transparent thermoplastic polymer, possesses moderate physical properties associated with low cost and composite systems based on this were studied in great detail [17-30]. It is reported that, for the PMMA+Al$_2$O$_3$ composite system, Al$_2$O$_3$ loading to the level of 45 vol %, increased the permittivity and has been attributed to the space charge polarization mechanism [18]. Similarly, permittivity as high as 19 has been reported for the PMMA composite comprising 40 vol % nanocrystallites of BaTiO$_3$ [19]. It is interesting to note frequency independent permittivity nature of poly(methyl methacrylate)/multi-walled carbon composite systems [20]. The PMMA based ternary composites also exhibited improved dielectric properties [21]. Apart from these, the introduction of fillers into the PMMA matrix also improved the thermal and mechanical properties [22,23]. These results suggest that one could chose PMMA combined with giant permittivity ceramics such as CCTO for the fabrication of composites with high permittivity. As CCTO possessing giant permittivity ($\varepsilon \sim 10^{4-5}$) which is nearly independent of frequency (upto 10 MHz) and low thermal coefficient of permittivity (TCK) over 100-600K temperature range [9], we thought it is worth investigating into two phase composite consisting of PMMA/CCTO, a heterogeneous system that would give rise to excellent thermal and dielectric properties for the energy storage devices. In this paper, we report the details pertaining to the fabrication and characterization of PMMA/CCTO composite by melt mixing and hot pressing technique.

**EXPERIMENTAL**

The solid-state reaction route was adopted for synthesizing CCTO ceramic powders [9,15]. The stoichiometric amounts of AR grade $CaCO_3$, CuO and $TiO_2$ was weighed, mixed using acetone and ball milled (300 rpm) for 5h. The homogeneous mixture thus obtained was dried in an electric oven for about 1h. This stoichiometric mixture was taken in a re-crystallized alumina crucible and heated at $1000^oC$ for 10h to obtain phase pure CCTO [9]. In order to get submicron particles, the CCTO powders were ball milled for about 12h using a planetary mill. PMMA, having Molecular weight of 1,10,000, (Make: LG Corporation) was used as matrix material. For the fabrication of composites, initially, the as received PMMA granules were heated at $210^oC$ in Brabender Plasticorder (Model:PLE331) till the PMMA granules are thoroughly melted. To this melt, CCTO powder (0 to 38 % by volume) was slowly added and mixed for 20 min at this temperature. The mixture was taken out from the Plasticorder and hot-pressed at this temperature to obtain a sheet of 100 $mm^2$ with 1.0 mm in thickness. A series of CCTO/PMMA composites were fabricated. The composite become brittle when the ceramic loading was beyond 38 % by volume. Hence, to obtain flexible composite which could be made into a variety of shapes, the ceramic loading has been restricted to maximum of 38 Vol %. Fig. 1 shows the flowchart depicting various steps involved in the fabrication of PMMA-CCTO Composites.

To examine the structure, an XPERT-PRO Diffractometer (Philips, Netherlands) was used. Scanning electron microscope (FEI-Technai TEM-Sirion) was used for the microstructure analysis of the composite samples. Thermo gravimetric (TGA) analyses were done using the TA Instruments (UK, Model: TGA Q500) in Nitrogen atmosphere at a flow rate of 60ml/min, and at a heating rate of 10 deg/min. For carrying out the dielectric measurements, the composite sheets were polished using fine emery paper to achieve smooth and parallel surfaces. Further, these samples were cleaned under ultra-sonication and subsequently the surfaces were electroded with silver paste and cured at $50^oC$. An LCR meter (Model: HP4194A) was used for the capacitance measurements as a function of frequency (100Hz–1MHz) at room temperature. The measurement accuracy of the instrument is less than 5%. The dielectric constant was calculated using the relation, $\varepsilon_r = C \times d / \varepsilon_o A$ where $C$ =capacitance, $d$ is the thickness of the sample, $\varepsilon_o = 8.854 \text{X} 10^{-12}$ F/m and A is the electrode area of the sample.

## 3.0 RESULTS AND DISCUSSION

### *3.1. X-Ray Diffraction studies:*

The X-ray diffraction patterns were recorded on the CCTO ceramic powders, as received PMMA and on the composites that were fabricated. X-ray diffraction pattern revealed that the as received PMMA is semicrystalline in nature (Fig.2(a)). The XRD data obtained for the as prepared CCTO powders (Fig 2(b)) are compared well with the ICDD data (01-075-1149) shown in Fig 2(c) and with that reported earlier [15]. The X-ray diffraction pattern recorded for the PMMA comprising 6 Vol % CCTO revealed its composite nature. In the case of PMMA with 38 vol % CCTO composite (Fig.2(e)), XRD pattern revealed only CCTO peaks as the intensities of these are dominants.

### *3.2. Thermal studies (TGA):*

In order to examine the thermal stability, thermal analyses were carried out on PMMA+CCTO composites as well as on pure PMMA for comparison. The TGA data obtained for the pure PMMA and the composites are illustrated in Fig.3. It has been observed that there is a change in the thermal degradation behaviour of PMMA with CCTO. The onset of decomposition temperature (temperature at 10% weight loss) [26] was found to increase for all the composites under study. The decomposition temperature onset accompanied by 10% weight loss for PMMA+6 vol % CCTO is $358.4^{o}$C and for the PMMA+38 vol % CCTO, it is $374.1^{o}$C, while for pure PMMA, it is $353.5^{o}$C. The onset of decomposition temperature is higher by $20^{o}$C than that of virgin PMMA. All the samples (PMMA and composites) indicated that there is no weight loss upto $270^{o}$C and thereafter, the degradation begins. Overall, the composites have better thermal stability than that of PMMA.

### *3.3. Microstructural Studies :*

Fig.4(a-d) illustrates the SEM micrographs recorded for the as prepared composites containing 6, 10, 21 & 38 Vol % of CCTO. The micrographs were recorded on the hot pressed composites (as prepared) and it is seen that (Fig.4 (a)) the crystallites are randomly distributed for PMMA+CCTO-6 Vol % composite and in the case of PMMA+CCTO-10 Vol % composite (Fig.4(b)) the crystallites are distributed evenly with less agglomeration accompanied by minimum porosity. In the case of PMMA+CCTO-38 Vol % composite (Fig.4(d)), the crystallites are well distributed and the porosity is almost negligible implying

that the fabrication of composites with good dispersion of ceramic particles into the polymer matrix has been achieved. The composite becomes brittle when the ceramic loading is increased beyond 38 Vol %.

### 3.4 Frequency dependence of room temperature permittivity

The room temperature effective permittivity data ($\varepsilon_{eff}$) for PMMA/CCTO composites for different volume percents of CCTO are given in fig.5a. The permittivity of the pure PMMA is around 4.9 @100Hz, which is nearly constant over the entire frequency range covered in the present investigation (fig.5a). As expected, the permittivity increases as the ceramic loading increases in the polymer matrix at all the frequencies under study. The permittivity has increased to 15.7@100Hz when the ceramic loading is increased to 38 Vol % in PMMA. The value of permittivity obtained for the PMMA+38 Vol % CCTO is higher than that of the pure PMMA and much lower than that of CCTO ceramic (fig.5a). The room temperature dielectric loss (D) recorded as a function of frequency is shown in fig.5.(b). The dielectric loss increased as the CCTO content increased in PMMA, but decreased with the increase in frequency. Around 10 kHz, there is a sudden drop in the loss value. It is known that the addition of fillers induces structural changes in the PMMA matrix [18] which may result in a sudden drop in the dielectric loss in the composites. The higher dielectric loss especially at low frequencies, is attributed to interfacial polarization/MWS effect [16,18]. The dielectric loss for all the composites lies below 0.1 for the entire frequency range under investigation. The dielectric loss obtained for pure PMMA is almost independent of the frequency (100Hz to 100MHz). In this work, it is to be noted that the dielectric loss obtained in CCTO/PMMA composites is remarkably low. For instance, the loss value obtained @100 Hz for PMMA+38 Vol % CCTO composite is around 0.094 and it has decreased to 0.011 @100MHz, demonstrating that this PMMA+CCTO composite may be exploited in the design and fabrication of capacitors for high frequency applications.

### 3.5 Temperature dependent dielectric properties

Figure.6(a&b) illustrates the temperature dependence of dielectric properties of PMMA+CCTO-38 vol % composite. Fig.6a shows the frequency dependent permittivity at different temperatures for the composites. As expected, the permittivity increased with increase in temperature (30°C-150°C temperature range) while it decreases as the frequency is increased from 100Hz to 100MHz. The permittivity value at 100 Hz is about 15.7 at room temperature, which has increased to about 26 at 150°C. Higher permittivity values at low frequency can be explained by invoking interfacial polarization mechanism, akin to that reported for the other composite system in the literature [14-18]. When the viscosity of the glass abruptly decreases in the vicinity of the glass-transition region, ions easily respond to the external electric field and the dielectric constant increases. The addition of CCTO has increased the permittivity of PMMA; however, the composite's permittivity is much smaller than that of the CCTO ceramic, which is a general feature associated with the polymer-ceramic composite systems. The frequency dependent dielectric loss at various temperatures is given in fig.6(b). It is observed that there is a relaxation at low frequency and is shifted to higher frequency as the temperature is increased. The shift is more noticeable at higher temperatures (beyond 100°C). The temperature dependent relaxations in PMMA has been well studied [28-30] and is reported that the PMMA nanocomposites exhibited three relaxations, $\gamma$, $\beta$ and $\alpha$ respectively [28]. We have noticed two relaxations, one at low frequencies and the other at the high frequencies, which are distinctly evident at the higher temperatures. It is also noticed that as the temperature increases, the relaxation peak and amplitude shifts towards higher frequencies. This might be due to the merging of both $\beta$ and $\alpha$ relaxation into a single dielectric dispersion peak at high temperatures [28-30].

The effective permittivity of polymer/filler composite material is dependent not only on the permittivity of the polymer matrix and the filler, size and shape of the filler and its volume fraction, but also on the permittivity of the interphase region and its volume. Hence, it is necessary to use the available models to predict the permittivity by combining the theory and the experiment. Various models have been developed for the 0-3 composites [31-34].

Fig.7 gives the room temperature (300K) permittivity of the composite at 10kHz (experimental) for different volume fractions of CCTO. The experimental value has been compared with that obtained using different models and is included in the same figure for comparison. According to the Maxwell's model for the diphasic mixture [31], the effective permittivity ($\varepsilon_{eff}$) of the composite is given by

$$\varepsilon_{eff} = \left( \frac{\delta_p \varepsilon_p \left( \frac{2}{3} + \frac{\varepsilon_c}{3\varepsilon_p} \right) + \delta_c \varepsilon_c}{\delta_p \left( \frac{2}{3} + \frac{\varepsilon_c}{3\varepsilon_p} \right) + \delta_c} \right) \qquad (1)$$

where, $\varepsilon_c$, $\varepsilon_p$, $\delta_c$ and $\delta_p$ are the permittivity of CCTO, PMMA, the volume fraction of the dispersoid and the polymer, respectively. The values that are substituted for $\varepsilon_c$ and $\varepsilon_p$ are 2500 and 4.05 (at 10kHz). It has been observed that the values obtained for $\varepsilon_{eff}$ by using the above expression deviates from the experimental value for all the volume fractions of CCTO under study.

The Clausius-Mossotti [32] was also used for predicting the effective permittivity ($\varepsilon_{eff}$) of a mixture of dielectrics composed of spherical crystallites dispersed in a medium. The effective permittivity ($\varepsilon_{eff}$) of the composite is calculated using the equation:

$$\varepsilon_{eff} = \varepsilon_p \left[ 1 + 3\delta_c \left( \frac{(\varepsilon_c - \varepsilon_p)}{\varepsilon_c + 2\varepsilon_p} \right) \right] \qquad (2)$$

The CCTO particles are non-spherical in nature and hence, the values obtained for $\varepsilon_{eff}$ by using the above expression deviates from the experimental value for all the volume fractions of CCTO under study.

Since the morphology of the crystallites play a definite role, taking into account the geometry of the crystallites, the effective medium theory (EMT) [33] and Yamada [34] models have been invoked. According to the EMT model,

$$\varepsilon_{eff} = \varepsilon_p \left[ 1 + \frac{f_c(\varepsilon_c - \varepsilon_p)}{\varepsilon_p + n(1 - f_c)(\varepsilon_c - \varepsilon_p)} \right] \qquad (3)$$

where $f_c$, is the volume fraction of the ceramic dispersed, $\varepsilon_c$, $\varepsilon_p$ and $n$ are the permittivity of the crystallites, polymer and the ceramic morphology fitting factor respectively. It is observed that the values obtained for $\varepsilon_{eff}$ by using the above expression is in close agreement with the experimental value for $f_{ccto}$ of 0.15 to 0.28. The $\varepsilon_{eff}$ value for $f_{ccto} = 0.15$, 0.21 & 0.28 is 7.2, 8.8 & 11.2 respectively, which are comparable with the experimental ($\varepsilon_{eff}$) value (7.7, 9.1, 11.6). The $\varepsilon_{eff}$ value obtained for $f_{ccto}$ of 0.10 & 0.40 deviates much from the experimental value.

The model that was developed by Yamada [34], has been employed to predict $\varepsilon_{eff}$ of the present composites. According to which,

$$\varepsilon_{eff} = \varepsilon_1 \left[ 1 + \frac{n f_{CCTO}(\varepsilon_2 - \varepsilon_1)}{n\varepsilon_1 + (\varepsilon_2 - \varepsilon_1)(1 - f_{CCTO})} \right] \qquad (4)$$

Here, $\varepsilon_1$ and $\varepsilon_2$ are, respectively, the permittivity of the polymer and ceramic, "$n$" is the parameter related to the geometry of the ceramic particles and $f_{ccto}$ is the volume fraction of the ceramic in the matrix. When the parameter "$n$" is around 5.0, the $\varepsilon_{eff}$ values for $f_{ccto}$ = 0.15, 0.21 and 0.28 is 7.5, 9.2 & 11.9 respectively, which are comparable with the experimental $\varepsilon_{eff}$ value (7.7, 9.1, 11.6). The $\varepsilon_{eff}$ value obtained for $f_{ccto}$ = 0.10 & 0.40 is at variance with the experimental value. These results are in agreement with the results obtained using the EMT model in this work. These data strongly suggests that the morphology of the crystallites is crucial in modelling the dielectric properties of the present composites.

To further rationalise the temperature dependence of relaxation processes, electrical modulus approach has been adopted. The use of electric modulus approach also helps in gaining an insight into the bulk response of materials. This would facilitate to circumvent the problems caused by electrical conduction which might mask the real dielectric relaxation process [35]. The complex electric modulus ($M^*$) is defined in terms of the complex dielectric constant ($\varepsilon^*$) and is represented as

$$M^* = (\varepsilon^*)^{-1}, \tag{5}$$

$$M^* = M' + iM'' = \frac{\varepsilon_r'}{(\varepsilon_r')^2 + (\varepsilon_r'')^2} + i\frac{\varepsilon_r''}{(\varepsilon_r')^2 + (\varepsilon_r'')^2}, \tag{6}$$

where, $M'$, $M''$, $\varepsilon'$ and $\varepsilon''$ are the real and imaginary parts of the electric modulus and dielectric constants, respectively. The real and imaginary parts of the electric modulus are calculated using Eq.(5) and are depicted in Fig.8 & 9 respectively. The $M'$ value obtained for both pure PMMA and composites (fig.8) lies below 0.21. The low frequency $M'$ values for the composite decreased when the CCTO loading is increased in the PMMA. For all the samples, the reduction in $M'$ indicates an increase in charge carrier density due to CCTO as well as polymer segmental motion [36]. The increase of CCTO in PMMA results in lower values of $M'$, implying that the real part of permittivity increases with ceramic filler [16] and as reported [36,37], negligible contributions arising from the electrode polarization. The Fig.9 shows the dispersion of $M''$ in the lower frequency region which is similar to the results that were reported in the literature and the asymmetric shape associated with this has been attributed to the non-Debye process [37]. Figure 10 shows the variation of imaginary part of electrical modulus ($M''$) as a function of frequency at various temperatures for the PMMA+CCTO (38 vol %) composite. As discussed earlier, the relaxation peak at lower frequency shifts to higher frequency as the temperature increases and merges into a single dielectric dispersion peak [28-30]. This is influenced by the interfacial polarization effect [38], as the charge accumulation around the ceramic particles shifts the relaxation peak to higher frequencies. This type of relaxation process is ascribed to the filler concentration [38].

The AC conductivity $\sigma$ *is* derived from the dielectric data using the equation (7):

$$\sigma^{'} = \varepsilon_{o}\omega\varepsilon^{''}$$

(7)

where $\varepsilon_{o}$ = 8.85.10$^{-12}$ F/m is the permittivity of the free space and $\omega = 2\pi f$ the angular frequency. It has been observed that the variation of AC conductivity as a function of frequency (not shown here) increases with the frequency for both the PMMA and the composites and as reported [18], the increase in AC conductivity at low frequency is due to the interfacial polarization and in the high frequency region, it is attributed to the electronic polarization.

**CONCLUSIONS**

The PMMA-CCTO composites exhibited better thermal stability than that of the pure PMMA as expected, the permittivity of PMMA increased with increase in CCTO content. The low dielectric loss that was exhibited by PMMA-CCTO composites could be exploited for high frequency capacitors applications. The shift in relaxation peak towards higher frequencies is ascribed to the merging of both β and α relaxations into a single dielectric dispersion peak. The increases in AC conductivity in the high frequency region is attributed to the electronic polarization associated with the composite.

**ACKNOWLEDGEMENT**

The management of Central Power Research Institute is acknowledged for the financial support (CPRI Project No. R-DMD-01/1415).

**REFERENCES**

[1]    Y.Bai, Z-Y.Cheng, V. Bharti, H.Xu, Q. M. Zhang, *J. Appl. Phys. Lett*, **76**, 3804 (2000).

[2]    R. Gregorio, M. Cestari, F. E. Bernardino, *J. Mater. Sci.*, **31**, 2925 (1996).

[3]    S. U. Adikary, H. L. W. Chan, C. L. Choy, B. Sundaravel, I. H. Wilson, *Compos. Sci. Technol.*, **62**, 2161 (2002).

[4]    F. Wen, Z. Xu, W. Xia, X. Wei, Z. Zhang, *Polym. Eng. Sci.*, **53**, (2012). DOI: 10.1002/pen.23312

[5]    Z. M. Dang, Y. H. Lin, C.W Nan, *Adv. Mater.*, **15**, 1625 (2003).

[6]    P. Kim, S.C. Jones, P.J. Hotchkiss, J.N. Haddock, B. Kippelen, S.R. Marder, J.W. Perry, *Adv. Mater.*, **19**, 1001 (2007).

[7]    C. Muralidhar, P. K. C. Pillai, *J. Mater. Sci.*, **23**, 1071 (1988).

[8]    S. L. Swartz, *IEEE Trans on.*, **25**, 935 (1990).

[9]    M. A. Subramanian, D. Li, N. Duran, B. A. Reisner, A. W. Sleight, *J. Solid State Chem.*, **151**, 323 (2000).

[10]   M. Arbatti, X. Shan, Z-Y. Cheng, *Adv. Mater.*, **19**, 1369 (2007).

[11]   B. Shri Prakash, K. B. R. Varma, *Compo. Sci. Technol.*, **67**, 2363 (2007).

[12]   E. Tuncer, I. Sauers, D. R. James, A. R. Ellis, M. P. Paranthaman, A. T. Tolga, S. Sathyamurthy, L. M. Karren, J. Li, A. Goyal, *Nanotechnol*, **18**, 25703 (2007).

[13]   F. Amaral, C. P. L. Rubinger, F. Henry, L. C. Costa, M. A. Valente, A. Barros-Timmons, *J. Non Crystalline Solids.*, **354**, 5321 (2008).

[14]   L. A. Ramajo, M. A. Ramírez, P. R. Bueno, M. M. Reboredo, M. S. Castro, *Mat. Research*, **11**, 85 (2008).

[15]   P. Thomas, K. T. Varughese, K. Dwarakanatha, K. B. R. Varma, *Compos. Sci. Technol.*, **70**, 539 (2010).

[16]   P.Thomas, S. Satapathy, K. Dwarakanatha, K. B. R. Varma, *eXPRESS. Polym. Lett.*, **4**, 632 (2010).

[17]   P. Thomas, R.S. Ernest Ravindran, K.B.R. Varma, IEEE 10th ICPADM: 2012 Jul 24-28 2012, Bangalore, India.

[18]   Bahaa Hussien, *Eur. J. Sci. Res.*, **52**, 236 (2011).

[19]   Y. Kobayashi, A. Kurosawa, D. Nagao, M. Konno, *Polym. Eng. Sci.*, **49**, 1069 (2009).

[20]   S. Kumar, T. Rath, B. B. Khatua, A. K. Dhibar, C. K. Das, *J. Nanosci Nanotechnol.*, **9**, 4644 (2009).

[21]   E.A. Stefanescu, X. Tan, Z. Lin, N. Bowler, M. R. Kessler, *Polym*, **51**, 5823 (2010).

[22]    T. E. Motaung, A. S. Luyt, M. L. Saladino, D. C. Martino, E. Caponetti, *eXPRESS. Polym. Lett*., **6**, 871 (2012).

[23]    H. Haitao Wang, S. Meng, P. Xu, W. Zhong, Q. Du, *Polym. Eng. Sci*., **47,** 302 (2007).

[24]    Wang Wen-Ping, Pan Cai-Yuan, *Polym. Eng. Sci.*, **44**, 2335 (2004).

[25]    Narayan Ch. Das, Yayong Liu, Kaikun Yang, Weiqun Peng, Spandan Maiti, Howard Wang, *Poly. Eng. Sci.*, **49**, 1627 (2009).

[26]    Marius C. Costache, Dongyan Wang, Matthew J. Heidecker, E. Manias, Charles A. Wilkie, *Polym. Adv. Technolo.,* **17**, 272 (2006).

[27]    J. Biros, T. Larina, J. Trekoval, J. Pouchly, *Colloi. Polym. Sci.,* **260**, 27 (1982).

[28]    A. C. Comer, A. L. Heilman, D. S. Kalika, *Polym*., **51**, 5245 (2010).

[29]    J. L. G. Ribelles, R. D. Callwa, *Polym Phys*., **23**, 1297 (1985).

[30]    R. Bergman, F. Alvarez, A. Alegria, J. Colmenero, *J.Non-Crystalline Solids*, **235-237**, 580 (1998).

[31]    A. L. Efros, B. I. Shklovskii, *Phys. Stat. Solidi B*, **76**, 475 (1976).

[32]    D. M. Grannan, J. C. Garland, D. B. Tanner, *Phys. Rev. Lett*., **46**, 375 (1981).

[33]    Y. Song, T. W. Noh, S-I. Lee, J. R. Gaines, *Phys. Rev. B*, **33**, 904 (1986).

[34]    C. Pecharroman, J. S. Moya, *Adv. Mat.*, **12**, 294 (2000).

[35]    V. Provenzano, L. P. Boesch, V. Volterra, C. T. Moynihan, P. B. Macedo, *J. Ame. Ceram. Soc.*, **55**, 492 (1972).

[36]    P. Sharma, D. K. Kanchan, N. Gondaliya, *Open J. Organic Polym Mater.*, **2**, 38 (2012).

[37]    A. K. Tomar, S. Mahendia, P. R. Chahal, S. Kumar, *Synthetic Metals*, **162**, 820-826 (2012).

[38]    G. A. Kontos, A. L. Soulintzis, P. K. Karahaliou, G. C. Psarras, S. N. Georga, C. A. Krontiras, M. N. Pisanias, *eXPRESS. Polym. Lett*, **1**, 781 (2007).

## *FIGURE CAPTIONS*

Figure.1. Flowchart depicting various steps involved in the fabrication of PMMA-CCTO composite

Figure.2. X-ray diffraction patterns for (a) pure PMMA, (b) $1000^{o}C/10h$ –Phase-pure CCTO, (c) CCTO-JCPDS, (d) PMMA+6 Vol % CCTO and (e) PMMA+38 Vol % CCTO.

Figure.3. Thermal analysis (TG ) for the (a) pure PMMA , (b) 6, (c) 10, (d) 21, (e) 28 and (f) for 38 Vol % PMMA-CCTO composite.

Figure.4. Scanning electron micrographs of CCTO/PMMA composite for different volume percent of CCTO (a) 6, (b) 10, (c) 21 and (d) for 38 Vol % PMMA-CCTO composite.

Figure.5. Frequency dependent (a) dielectric permittivity and (b) dielectric loss measured at room temperature (300K) for different Vol % PMMA-CCTO composite.

Figure.6. Frequency dependent (a) dielectric permittivity and (b) dielectric loss measured at various temperatures for PMMA-CCTO- 38 % composite.

Figure.7. Variation of effective dielectric constant ($\varepsilon_{eff}$) (measured at 300K and 10kHz of PMMA-CCTO composite as a function of volume fraction of CCTO particles ($f_{CCTO}$). For comparison, the calculations by using various models are also shown.

Figure.8. Real ($M'$) parts of electric modulus as a function ffrequency for different volume percents of CCTO.

Figure.9. Imaginary ($M''$) parts of electric modulus as a function ffrequency for different volume percents of CCTO.

Figure.10. Electric modulus spectra for PMMA+CCTO-38% composite at various temperatures as a function of frequency.

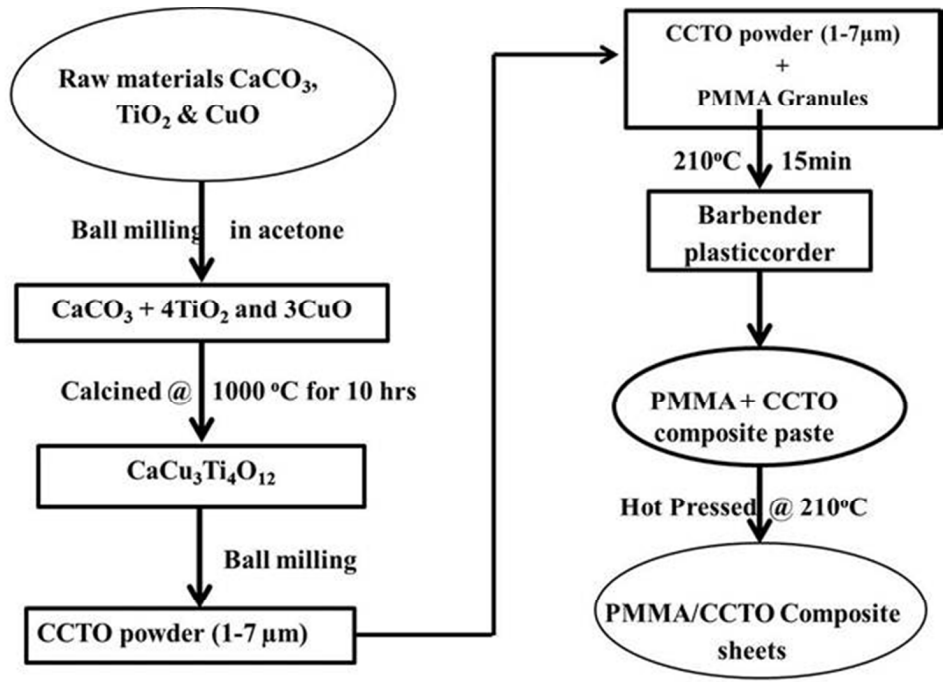

Figure.1Flowchart depicting various steps involved in the fabrication of  PMMA-CCTO composite
114x80mm (150 x 150 DPI)

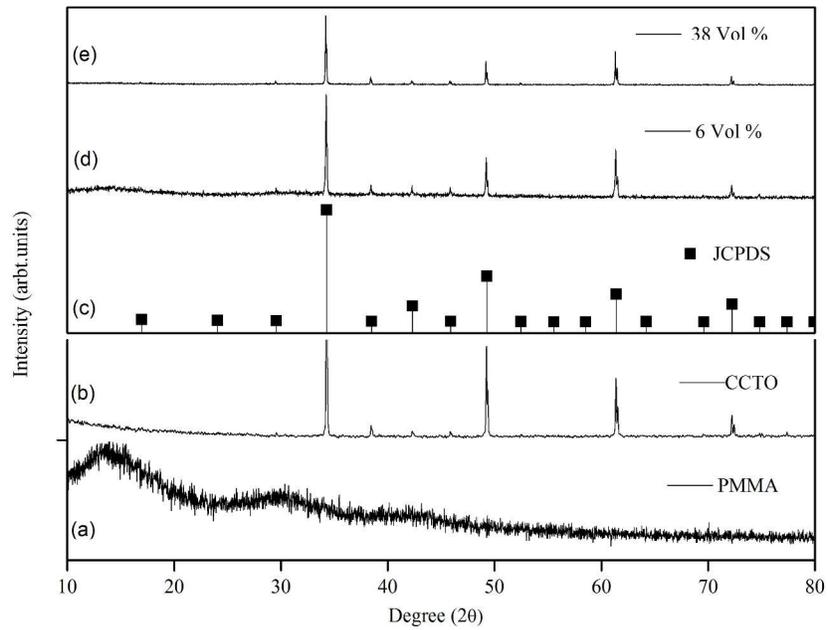

Figure.2. X-ray diffraction patterns for (a)  pure PMMA,  (b) 1000oC/10h –Phase-pure CCTO, (c) CCTO-
JCPDS, (d) PMMA+6 Vol % CCTO and (e) PMMA+38 Vol % CCTO.
287x201mm (300 x 300 DPI)

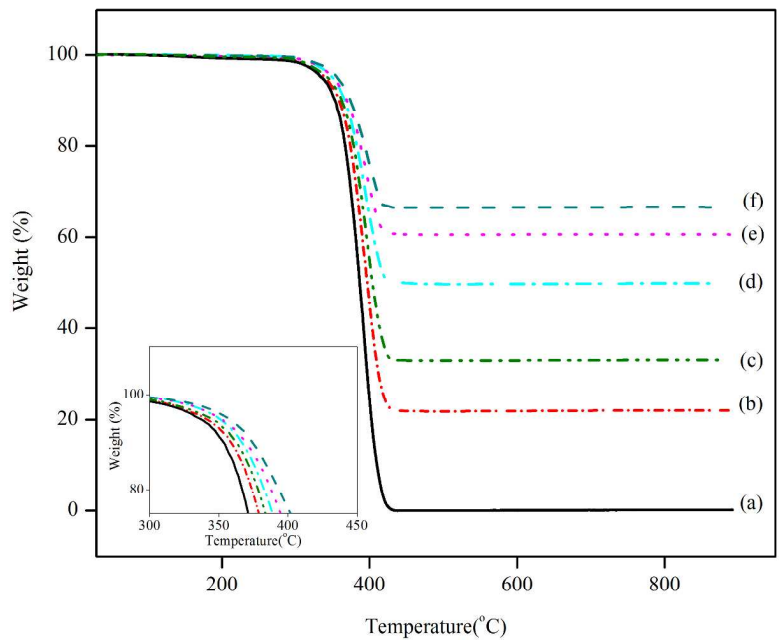

Figure.3. Thermal analysis (TG ) for the (a) pure PMMA , (b) 6, (c) 10, (d) 21, (e) 28 and (f) for 38 Vol % PMMA-CCTO composite.
271x207mm (300 x 300 DPI)

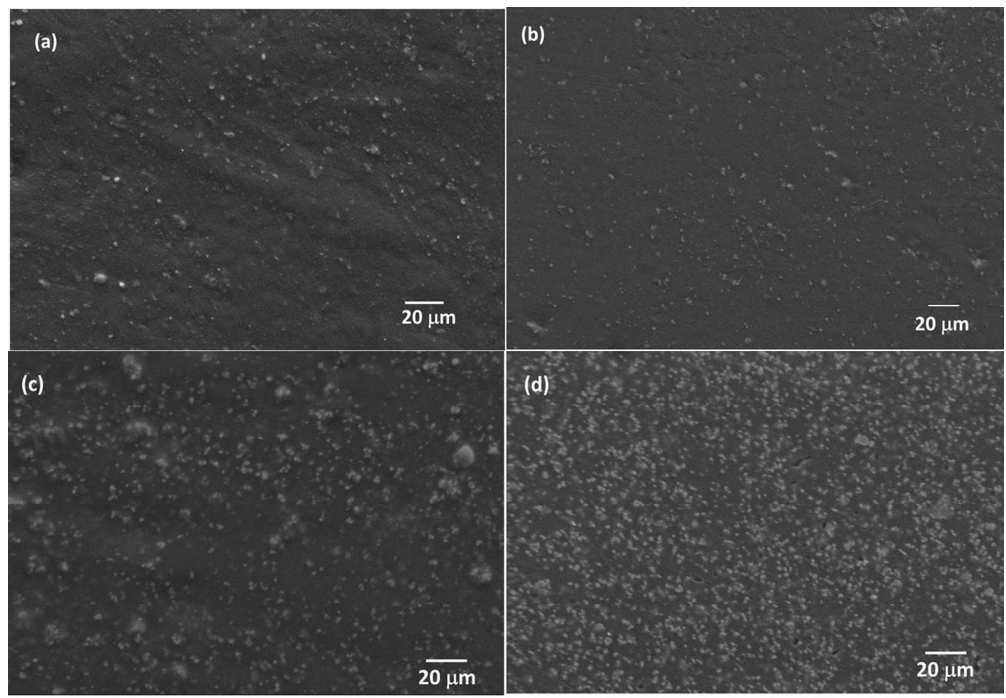

Figure.4. Scanning electron micrographs of CCTO/PMMA composite for different volume percent of  CCTO  (a) 6, (b) 10,  (c) 21 and (d) for 38 Vol %  PMMA-CCTO composite.
220x154mm (150 x 150 DPI)

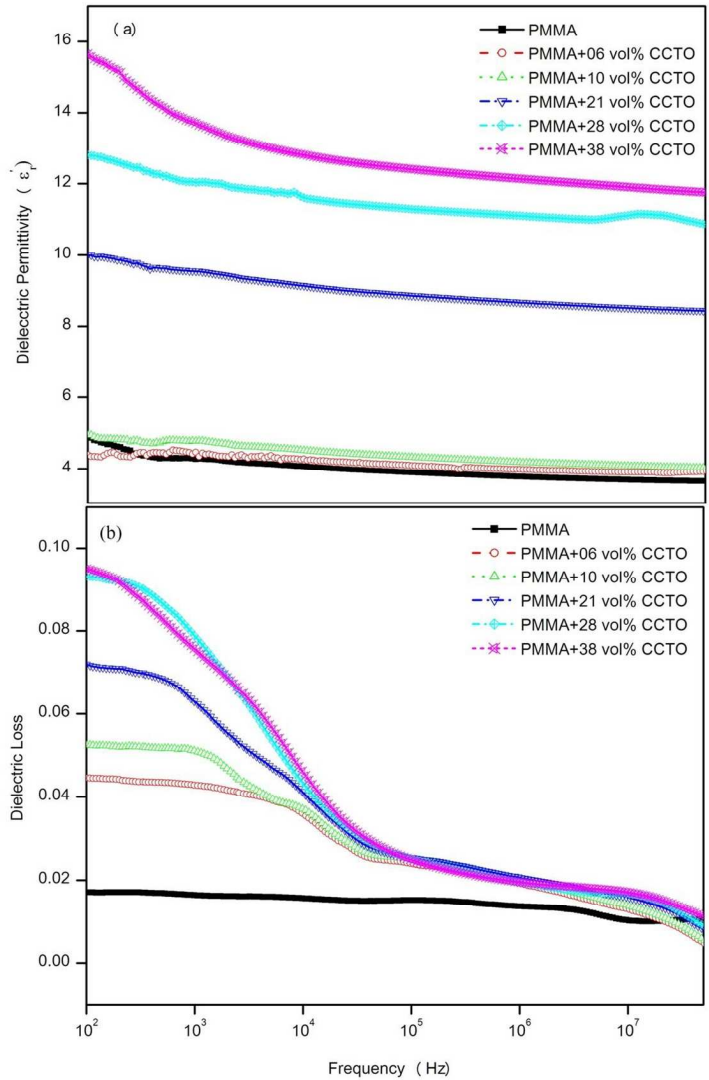

Figure.5. Frequency dependent (a) dielectric permittivity and (b) dielectric loss measured at room temperature (300K) for different Vol %  PMMA-CCTO composite.
248x328mm (150 x 150 DPI)

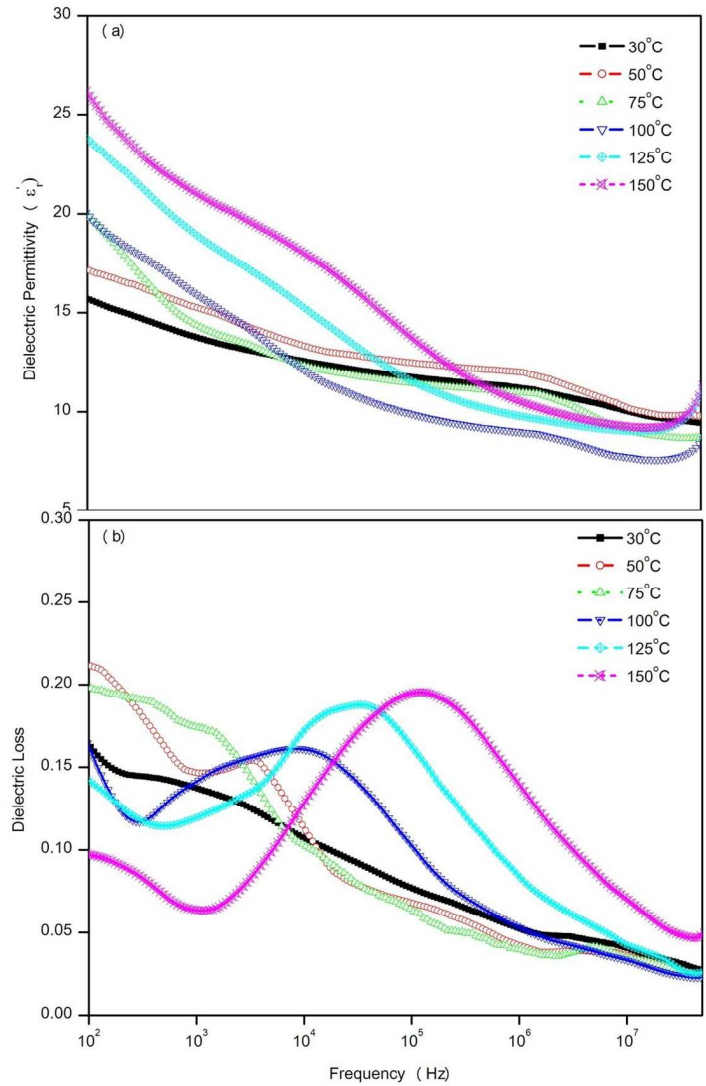

Figure.6 Frequency dependent (a) dielectric permittivity and (b) dielectric loss measured at various
temperatures for PMMA-CCTO- 38 % composite
248x329mm (150 x 150 DPI)

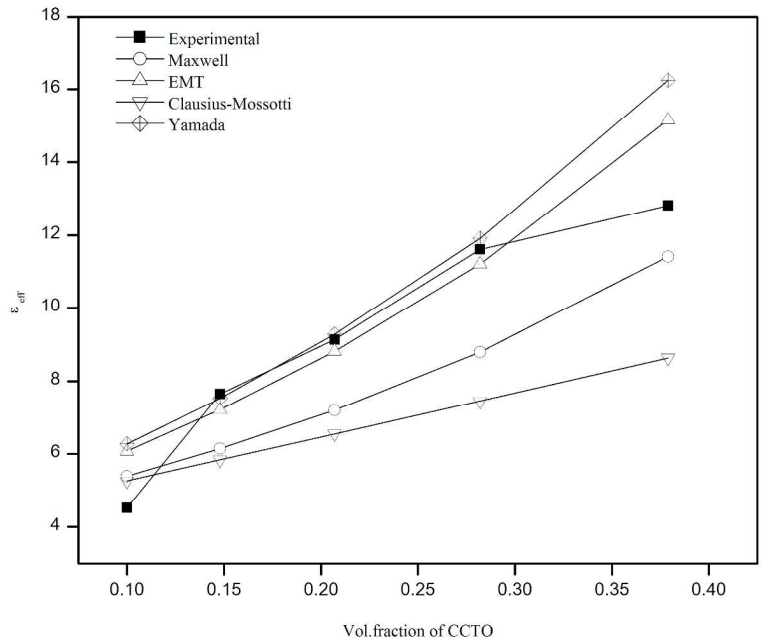

Figure.7. Variation of effective dielectric constant ($\varepsilon$eff) (measured at 300K and 10kHz of PMMA-CCTO composite as a function of   volume fraction of CCTO particles ($f$CCTO). For comparison, the calculations by using various models are also shown.
541x414mm (150 x 150 DPI)

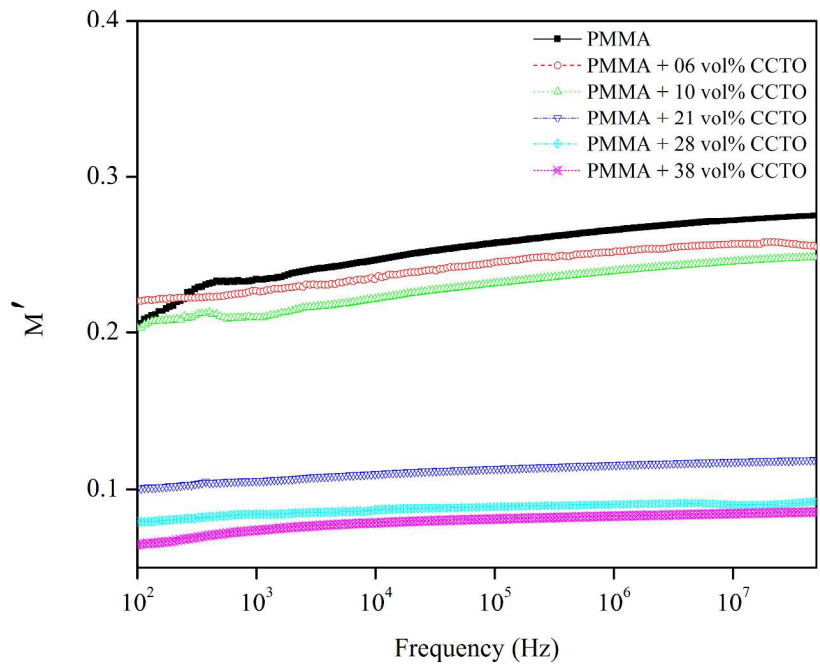

Figure.8. Real (M')  parts of electric modulus as a function ffrequency for different volume percents of CCTO.
271x207mm (300 x 300 DPI)

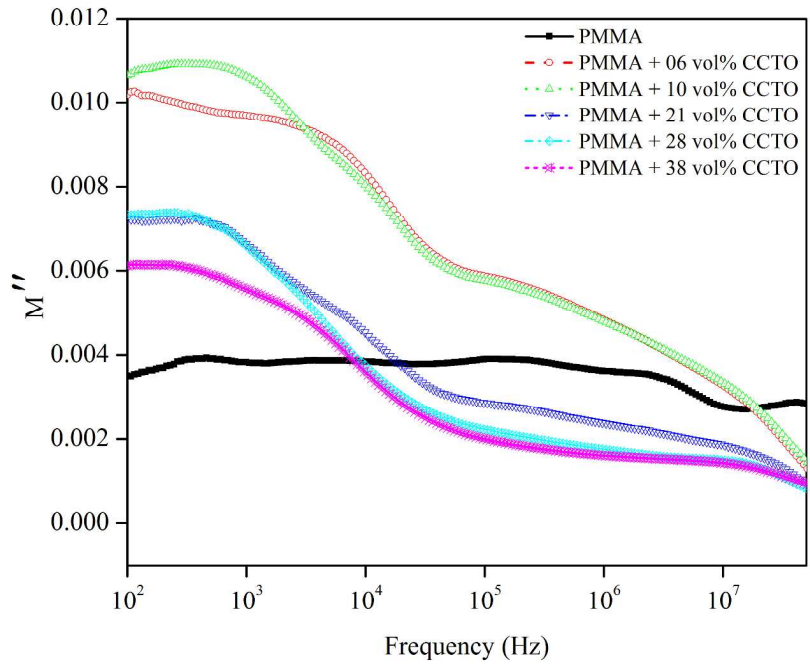

Figure.9. Imaginary (M'') parts of electric modulus as a function ffrequency for different volume percents of CCTO.
271x207mm (300 x 300 DPI)

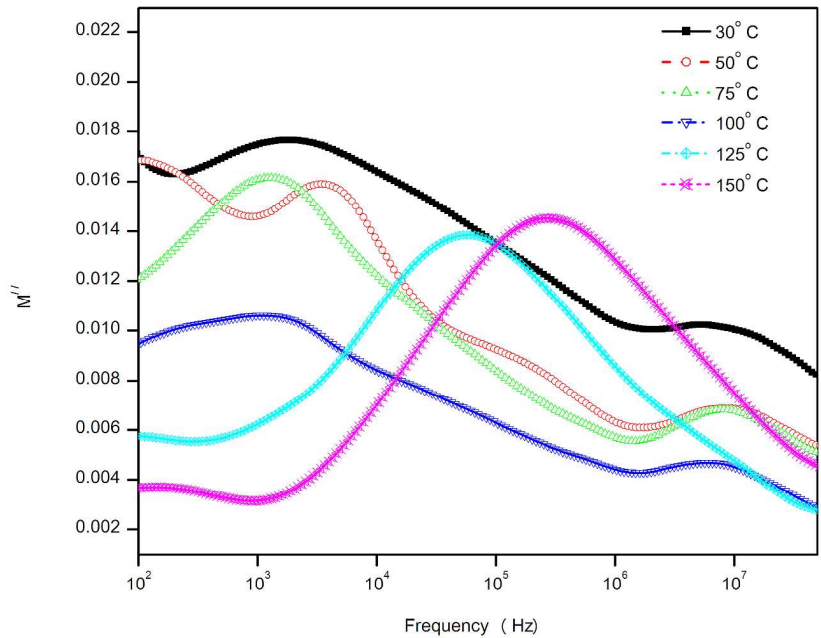

Figure.10. Electric modulus spectra for PMMA+CCTO-38% composite at various temperatures as a function of frequency
541x414mm (150 x 150 DPI)